\documentclass[twoside,11pt]{article}

\usepackage[cp1251]{inputenc}
\usepackage{graphicx}

\usepackage{amssymb}
\usepackage{amsmath}
\usepackage{amsfonts}
\usepackage{wasysym}

\begin{document}           

\title{\Large 
K-matter as Mach's principle realization
}
\author{V.E. Kuzmichev, V.V. Kuzmichev\\[0.5cm]
\itshape Bogolyubov Institute for Theoretical Physics,\\
\itshape National Academy of Sciences of Ukraine, Kiev, 03680 Ukraine}

\date{}

\maketitle

\begin{abstract}
It is shown that if one takes into account Mach’s principle in the form which follows from quantum theory and considers it as a complementary constraint between the parameters 
which characterize the energy density and geometry of the universe in addition to Einstein equations for
a FRW universe, non-relativistic matter transforms into an analogue of K-matter.
The exact solutions of the Einstein equations for the universe with such matter and cosmological constant are found. It is demonstrated that the Machian universe under consideration with a nonzero cosmological constant is
equivalent to the open de Sitter universe. In the limit of zero cosmological constant such a universe evolves as a Milne universe, but in contrast to it, it contains matter with nonzero energy density. The possible application of proposed approach to the description of the present cosmological data is discussed. The problem of the age of the universe is considered as an example.
\end{abstract}


\section{Introduction}

As is well-known, the standard $\Lambda$CDM model gives the satisfactory description of the most
of the present cosmological data under the assumption of the existence of dark energy 
as the largest constituent of mass-energy in the universe.  
It is believed that a high level of fine-tuning is required
in this model. Even if the smallness of cosmological constant and ``coincidence problem'' 
(an almost equal contribution of matter and dark energy to the total energy budget of the universe at 
the present epoch) are not problems in themselves \cite{Bia},  nevertheless one must take into account 
the phenomenological character of the $\Lambda$CDM model, as regards the choice of the 
form of the energy density and equation of state.
It should not be ignored that there were some indications that specific cosmological observations
differed from the predictions of the $\Lambda$CDM model at statistically significant level \cite{Pe}.
It makes the search for alternative models not as unreasonable as it might seem.

It was noticed that the models (called ``coasting cosmology'', since in such models 
the universe expands with constant velocity \cite{Kol}) in which the scale factor $R$ of the universe 
depends on synchronous proper time $t$ linearly, $R(t) \sim t$, 
agrees good enough with the cosmological observations \cite{Dev,Ben}. In particular, this approach
does not suffer from the horizon problem. It is in concordance with the early universe nucleosynthesis
constraints and fits to type Ia supernovae data. One can expect that such a scenario 
will agree with estimates of age of the
universe in comparison to ages of old objects and provide the degree scale for the first acoustic peak of
the cosmological microwave background. It is compatible with constraints from large scale
structure formation and agrees with the physics of recombination as deduced from cosmic microwave
background anisotropy. 
There are also expectations that the primordial lithium problem (a discrepancy 
between the primordial lithium predicted from the WMAP data and the stellar abundance determinations) \cite{Coc} might be resolved under the assumption of a linear evolution of the scale factor \cite{Dev,Ben}.

The linear dependence of a scale factor on time can be associated with a Milne model of the universe.
This model is based on the assumptions that the universe is open ($k = -1$) 
and that the gravitational action of matter can be neglected (the energy density $\rho = 0$). 
It cannot be correct near the point of initial cosmological singularity, $t = 0$, since
in this limit the energy density of matter tends to infinity and gravity cannot be neglected. One attempt
to settle this problem was to consider a model of a universe (called ``Dirac - Milne'' universe by analogy of sea
of positive and negative energy states proposed by Dirac) 
containing equal quantities of matter
and antimatter under the assumption that antimatter is characterized by a negative gravitational mass
\cite{Ben}. 

The linear dependence of a scale factor $R$ on time can be achieved
in cosmological models in which  the energy density $\rho \sim R^{-2}$ and the total pressure 
$p = - \frac{1}{3} \rho$.
Such ``coasting cosmologies'' describe the universe dominated by exotic ``K-matter'' which may be related to cosmic strings 
\cite{Kol,Got}. 

In the present paper we show that if one takes into account an additional constraint between the parameters 
which characterize the energy density and geometry of the universe in addition to Einstein equations for
a Friedmann-Robertson-Walker (FRW) universe, 
non-relativistic matter with the energy density evolving as $\rho \sim R^{-3}$ transforms into an analogue of K-matter.
This constraint can be interpreted as Mach's principle \cite{Mach} in Sciama's formulation \cite{Sc1,Sc2}. 
Being introduced to explain the inertial forces acting on a body via the quantity and distribution of matter in the whole universe, nowadays Mach's principle has many definitions \cite{Bon,Bar}. Despite its simplified character, Sciama’s linearized theory gives a specific mathematical relation between the parameters of the universe instead of general statement of Mach's principle.

\section{Quantum roots of Sciama's relation}

Sciama's relation obtains a natural explanation in
the framework of quantum isotro\-pic cosmological model \cite{Ku1,Ku2}. Generally speaking, quantum 
theory adequately 
describes properties of various physical systems. Its universal validity demands that the universe as a 
whole must obey 
quantum laws as well. Since quantum effects are not a priori restricted to certain scales, then one should not conclude in advance 
that they cannot have any impact on processes at scales larger than Planckian 
(more detailed arguments can be found, e.g., in Refs. \cite{Mes}).

Quantum theory for a homogeneous and isotropic universe can be constructed on the basis of a 
Hamiltonian formalism with the use of material reference system as a dynamical system \cite{Ku1,Ku2}. 
Defining the time parameter or the ``clock'' variable, it is possible to pass from 
the Wheeler–DeWitt equation to the Schr\"odinger-type equation.
The similar equations containing a time variable defined by means of coordinate condition were 
considered by a number of authors under the quantization of the FRW universe 
(see, e.g., Refs. \cite{Lun}).
Using the Schr\"odinger-type equation one can obtain equations of motion for the expectation values of 
a scale factor and its conjugate momenta. These equations pass into the equations of general relativity 
when the dispersion around the expectation values for a scale factor, matter fields and their conjugate 
momenta can be neglected.

Under this approach, in semi-classical limit, the equations of the theory are reduced 
to the form of Einstein equations for the FRW universe \cite{Ku2}. Such a quantum theory predicts that the following relation must hold
for the expectation value of the scale factor $R$ in the state $| M \rangle$ which describes the universe with the definite total amount of mass $M$ much larger than Planck mass, $M \gg M_{P}$,
\begin{equation}\label{1}
\langle M | R | M \rangle = G M,
\end{equation}
$G$ is the Newtonian gravitational constant (for details, see Refs. \cite{Ku2}). In classical limit, it appears to be possible to pass from the expectation value $\langle M | R | M \rangle$ to the classical value of the scale factor $R(t)$ which evolves in time in accordance with the Einstein equations for the FRW universe
\begin{equation}\label{2}
\dot{R}^{2} = \frac{8\pi G}{3}\,\rho R^{2} + \frac{\Lambda}{3}\,R^{2} - k, \qquad
\ddot{R} = - \frac{4\pi}{3}(\rho + 3p)R + \frac{\Lambda}{3}\,R,
\end{equation}
where 
\begin{equation}\label{3}
    \rho = \frac{M}{(4\pi /3) R^{3}} 
\end{equation}
is the energy density of matter with the mass $M$ in the equivalent flat-space volume $(4\pi /3) R^{3}$, $\Lambda$ is the cosmological constant,
\begin{equation}\label{4}
    p = - \rho - \frac{R}{3}\,\frac{d\rho}{dR}
\end{equation}
is the isotropic pressure, and $k = +1,0,-1$ for spatially closed, flat or open models. 
In semi-classical limit, the relation (\ref{1}) takes the form of Sciama's inertial force law which
describes Mach's principle \cite{Sc1,Sc2},
\begin{equation}\label{5}
    R = G M.
\end{equation}

The same equality between the mass and ``radius'' of the universe was considered by Whitrow and Randall \cite{Whi}. It is also similar to the relation valid for the Einstein universe (see, e.g., Ref. \cite{Tol}).

For the present-day universe the radius of its observed part is estimated as $R_{0} \sim 10^{28}\mbox{cm} \sim 10^{61}$ (in units of Planck length $l_{P} \sim 10^{-33} \mbox{cm}$), the mass-energy is $M_{0} \sim 10^{56}\mbox{g} \sim 10^{80} \mbox{GeV}\sim 10^{61}$ (in units of Planck mass $m_{P} \sim 10^{19} \mbox{GeV}$), and the mean energy density equals to $\rho_{0} \sim 10^{-29}\mbox{g cm}^{-3} \sim 10^{-122}$ (in units of Planck energy density $\rho_{P}  \sim 10^{93} \mbox{g cm}^{-3}$). It means that nowadays $\rho_{0} \sim G^{-1} R_{0}^{-2}$. Then from the definition of energy density $\rho_{0} \sim M_{0}R_{0}^{-3}$, it follows that the relation $R_{0} \sim GM_{0}$ must hold. The same conclusion can be made from the exact equation (\ref{5}). Since this equation must be true for an arbitrary chosen instant of time $t$, there arises the problem of mass increase, as interpreted from the point of view classical cosmology. Namely, it follows that total mass increases proportionally to a scale factor, $M(t) \sim R(t)$, if the gravitational constant $G$ and velocity of light $c$ are both constant. This difficulty can be resolved, in particular, if one supposes that the natural constants $G$ or $c$ change with time. 

The questions arised in connection with these problems were discussed using different frameworks and for different purposes.
According to Dirac’s large number hypothesis, the Newtonian constant $G$ must depend on time, so that
$G \sim t^{-1}$ and $R \sim t^{1/3}$ \cite{Di1} or $G \sim t^{-1}$ and $R \sim t$ \cite{Di2}. 
In the Brans-Dicke theory the constant $G$ is related to the average value of some dynamical scalar field $\phi$ which is coupled to the mass density $\rho$ of the universe, $\langle \phi \rangle \approx G^{-1}$, where $\langle \phi \rangle \sim \rho R^{2}$ \cite{Bra,Wei}.
Models with varying speed of light were applied in order to solve the horizon, flatness,  cosmological constant, and other cosmological problems (see, e.g., Refs. \cite{Pet}). Matter creation processes in the context of the cosmological models and their influence on the evolution of the universe were studied in Refs. \cite{Lim}. 

If we go back and consider the equation (\ref{5}) as following from the relation (\ref{1}), then we can interpret it in terms of quantum theory. In quantum model the state vector of isotropic universe is a superposition of all possible $| M \rangle$ - states which are not orthogonal between themselves, so that the inner product $\langle M_{1} |M_{2} \rangle \neq 0$, and the universe can transit spontaneously from the state with the mass $M_{1}$ to the state with the mass $M_{2} \neq M_{1}$ with nonzero probability $P(1\rightarrow 2) = |\langle M_{1} |M_{2} \rangle|^{2}$. For example, the probability of transition of the universe from the ground state (with respect to gravitational field) to any other state obey the Poisson distribution with the mean number of occurrences $n = \frac{1}{2}(M_{2} - M_{1})^{2}$ (for more details, see Refs. \cite{Ku2}). Then $R_{1} \rightarrow R_{2}$, when $M_{1} \rightarrow M_{2}$. If one would try to interpret this result in terms of the Newtonian cosmology, describing the universe as a flat Euclidean 3-space filled with a uniform matter with the energy density $\rho (t)$ (\ref{3}), such a transition would correspond to the passage to the sphere of radius $R_{2} > R_{1}$  which includes a mass $M_{2} > M_{1}$.

\section{FRW equations with Mach's principle}

If one assumes that Mach's principle is a fundamental law of nature, it must be implemented into the 
classical field equations. One point of view is that Einstein's field equations need not to be modified,
while Mach's principle should be considered as an additional condition. Such an approach was chosen
by Wheeler who proposed to understand Mach's principle as a selection rule (boundary condition) of
the solutions of the field equations \cite{Whe}. The Brans-Dicke theory mentioned above 
uses another way in which the field equations are generalized to become Machian \cite{Bra,Hel}.

Since in our approach Mach's principle in the form (\ref{5}) follows from quantum theory in semi-classical
limit, under classical description it can be introduced as an addition constraint and added to 
the classical field equations (\ref{2}). With account of the constraint (\ref{5}), 
the energy density of matter (\ref{3}) takes the form of K-matter energy density with the corresponding
equation of state,
\begin{equation}\label{6}
    \rho = \frac{3}{G} \frac{1}{4 \pi R^{2}}, \qquad p = - \frac{1}{3} \rho.
\end{equation}
According to common classification (see, e.g. Ref. \cite{Dym}), 
matter with such an equation of state can be attributed to strings,
since it naturally appears in string cosmology. But in this approach it does not mean that the universe is string-dominated.
The energy density and pressure in the form (\ref{6}) arise as an effect of an additional constraint between the global geometry and the total amount of matter in the universe as a whole.

The field equations are reduced to the form
\begin{equation}\label{7}
\dot{R}^{2} = \frac{\Lambda}{3}\,R^{2} + (2 - k), \qquad
\ddot{R} = \frac{\Lambda}{3}\,R.
\end{equation}
Their solution is
\begin{equation}\label{8}
    R (t) = \sqrt{\frac{3 (2 - k)}{\Lambda}} \sinh \left(\sqrt{\frac{\Lambda}{3}} t \right), \qquad R (0) = 0.
\end{equation}
Expansion of this solution for small $|\frac{\Lambda}{3} t^{2}|$ yields
\begin{equation}\label{9}
    R (t) = \sqrt{2- k} t \left[1 + \frac{1}{6} \left(\frac{\Lambda}{3} t\right)^{2} + \ldots \right].
\end{equation}
From the Hubble expansion rate
\begin{equation}\label{10}
    H (t) = \frac{\dot{R}}{R} = \sqrt{\frac{\Lambda}{3}} \coth \left(\sqrt{\frac{\Lambda}{3}} t \right),
\end{equation}
one obtains the expansion in the same limit
\begin{equation}\label{11}
    H t = 1 + \frac{1}{3} \left(\frac{\Lambda}{3}\right) t^{2} - 
    \frac{1}{45} \left(\frac{\Lambda}{3}\right)^{3} t^{4} + \ldots
\end{equation}
If $\Lambda \neq 0$, the expressions for the scale factor (\ref{8}) and the Hubble expansion rate 
(\ref{10}) are equivalent to the respective expressions for the de Sitter model of the universe with 
$k = -1$.

In the limiting case $\Lambda = 0$ it appears that
\begin{equation}\label{12}
    R (t) = \sqrt{2- k} t, \qquad H t = 1.
\end{equation}
This solution formally coincides with the solution of Milne model of open universe ($k = -1$), 
$R (t) \sim t$.
But in contrast to the Milne model, where the energy density of matter vanishes, $\rho = 0$, in
the case under consideration the energy density of matter is nonzero,
\begin{equation}\label{13}
    \rho = \frac{3 H^{2}}{4 \pi G (2 - k)}.
\end{equation}
For a spatially flat universe ($k = 0$) this density equals to the critical density, 
$\rho = \rho_{c} \equiv \frac{3 H^{2}}{8 \pi G}$.

The equation (\ref{13}) can be rewritten in the Whitrow-Randall form \cite{Whi}, 
$$
  G \rho t^{2} = \frac{3}{4 \pi} \frac{1}{n},
$$
i.e. $G \rho t^{2}$ is an invariant determined by the parameter $n = 2 - k$ characterizing the geometry of the universe.

Introducing a dimensionless parameter $K$ as in the model of K-matter,
\begin{equation}\label{14}
    K \equiv \frac{8 \pi G}{3} \rho R^{2},
\end{equation}
and using (\ref{6}), one finds that $K = 2$. This value agrees with the observational constraints on 
the parameter $K$ obtained by Kolb \cite{Kol} and Gott and Rees \cite{Got}.

The calculations with the parameters for standard $\Lambda$CDM model give the same value of
$H_{0} t_{0}$ for the present-day universe as follows from Eq.~(\ref{12}). 
Really, using the WMAP 7-year data \cite{Lar} for the age of the universe $t_{0} = 13.75 \pm 0.13$ Gyr 
and the Hubble parameter $H_{0} = 71.0 \pm 2.5$ km s$^{-1}$ Mpc$^{-1}$, 
one finds: $H_{0} t_{0} = 0.998 \pm 0.045$.
At the same time, substituting the cosmological constant $\Lambda = (1.302 \pm 0.143) \times 10^{-56}$ cm$^{-2}$
which corresponds to the dark energy density parameter $\Omega_{\Lambda} = 0.734 \pm 0.029$
\cite{Lar} into Eq. (\ref{10}) with the corresponding age of the universe $t_{0}$, one gets a somewhat excessive value:
$H_{0} t_{0} = 1.233 \pm 0.029$. It is necessary to keep in mind, of course, that 
the use of the values of the parameters of $\Lambda$CDM model in these estimations of
$H_{0} t_{0}$ has only illustrative character, since Eqs. (\ref{8})-(\ref{11}) were obtained under the
different model assumptions.

In the model, where the scale factor depends on time linearly (\ref{12}), the age of the universe and
the Hubble expansion rate depend on the redshift $z$ according to the simple laws
\begin{equation}\label{15}
    t = \frac{1}{(1 + z) H_{0}}, \qquad H = H_{0} (1 + z).
\end{equation}
For the present expansion rate measured by Hubble Space Telescope observations,
$H_{0} = 73.8 \pm 2.4$ km s$^{-1}$ Mpc$^{-1}$ \cite{Rie}, the age of the universe appears to
be equal $t_{0} = 13.26 \pm 0.43$ Gyr. This value does not differ drastically from the value predicted by
the WMAP 7-year data for the $\Lambda$CDM model, and it lies within the expected limit of 12 to 14 Gyr.

\section{Conclusion remarks}

In the coasting cosmological models considered without reference to Mach's principle or matter creation,
it is assumed the existence of a specific form of matter, such as K-matter with the energy density, which 
decreases in expansion as $R^{-2}$, or such a matter, whose energy density can be neglected in the 
open universe (Milne universe). The incorporation of Mach's principle into the theory does not
change the physical properties of matter itself (such as a perfect fluid in the form of dust with the
corresponding equation of state), but it takes into account the constraint which reflects collective behavior 
of matter in the universe considered as a whole.  
The local properties of the matter are not affected by Mach's principle.

The horizon problem, the luminosity distance-redshift relation, the angular diameter distance-redshift
relation, and the galaxy number count as a function of redshift in the model of the FRW universe with
energy density $\rho \sim R^{-2}$ were studied by Kolb \cite{Kol}. In the case of a K-dominated universe, 
kinematic tests limit the parameter $K$ to be $K \gtrsim 1$. In the model which takes into
account Mach's principle in the form (\ref{5}) the universe behave as K-dominated with
the parameter $K = 2$ which agrees with the analysis of Refs. \cite{Kol,Got}.

There is some indication that in the cosmological model where the scale factor linearly depends on time, 
the light element abundances, the position of the first acoustic peak of the CMB can be satisfactorily described \cite{Dev,Ben}.

From the analysis of type Ia supernovae discovered by the Supernova Cosmology Project, it follows that
the data are consistent with the model in which the mass density and cosmological-constant energy 
density vanish, $(\Omega_{M},\Omega_{\Lambda}) = (0,0)$ \cite{Per}. It means that the model 
characterized by linear dependence of the scale factor on time agrees well with the SNe Ia observations 
\cite{Dev}. It was shown that the accelerating expansion of the present-day universe extracted from
the observed luminosity of the type Ia supernovae can be explained by the theory which takes into
account the feedback coupling between geometry and matter (Mach's principle) \cite{Ku3}.

In the model which accounts for Mach's principle, 
an assumption of large amounts of dark energy in the universe is not required to explain 
cosmological observations.
The cosmological model with the scale factor $R$ which evolves in time according to the equation 
(\ref{9}) with zero or small cosmological constant can be a good alternative to the standard cosmological 
model.


\begin{thebibliography}{99}
\itemsep -6pt plus 1pt minus 1pt
\bibitem{Bia} E. Bianchi and C. Rovelli, arXiv:1002.3966 [astro-ph.CO].

\bibitem{Pe} L. Perivolaropoulos, arXiv:1104.0539 [astro-ph.CO]; in: The Problems of Modern Cosmology, ed. P.M. Lavrov, Tomsk State Pedagogical University Press, Tomsk, 2009 [arXiv:0811.4684 [astro-ph]].

\bibitem{Kol} E.W. Kolb, Astrophys. J. 344 (1989) 543.

\bibitem{Dev} A. Dev, M. Safonova, D. Jain, and D. Lohiya, Phys. Lett. B 548 (2002) 12 [arXiv:astro-ph/0204150]; S. Gehlaut, P. Kumar Geetanjali, and D. Lohiya, arXiv:astro-ph/0306448 (2003); G. Sethi, A. Dev, and D. Jain, arXiv:astro-ph/0506255 (2005); A. Dev, D. Jain, and D. Lohiya, arXiv:0804.3491 [astro-ph] (2008); M. Sethi, A. Batra, and D. Lohiya, Phys. Rev. D60 (1999) 108301.

\bibitem{Ben} A. Benoit-Levy and G. Chardin, arXiv:0903.2446 [astro-ph.CO] (2009); arXiv:0811.2149 [astro-ph] (2008).

\bibitem{Coc} R.H. Cyburt, B.D. Fields, and K.A. Olive, Phys. Lett. B567 (2003) 227 [arXiv:astro-ph/0302431]; JCAP 0811 (2008) 012 [arXiv:0808.2818 [astro-ph]];
A.Coc, E. Vangioni-Flam, P. Descouvemont, A. Adahchour, and C. Angulo, Astrophys. J.  600 (2004) 544 [arXiv:astro-ph/0309480]; D.N. Friedel, A. Kemball, and B.D. Fields, arXiv:1106.2471 [astro-ph.CO].

\bibitem{Got} J.R. Gott and M.J. Rees, MNRAS 227 (1987) 453.

\bibitem{Mach} E. Mach, Die Mechanik in ihrer Entwickelung: historisch-kritisch dargestellt, F.A. Brockhaus, Leipzig, 1897.

\bibitem{Sc1} D.W. Sciama, MNRAS 113 (1953) 34.

\bibitem{Sc2} D.W. Sciama, Modern Cosmology, Cambridge University Press, Cambridge, 1971.

\bibitem{Bon} H. Bondi and J. Samuel, Phys. Lett. A 228 (1997) 121 [arXiv:gr-qc/9607009].

\bibitem{Bar} J. Barbour, Found. Phys. 40 (2010) 1263 [arXiv:1007.3368 [gr-qc]].

\bibitem{Ku1} V.V. Kuzmichev, Ukr. J. Phys. 43 (1998) 896; Phys. Atom. Nucl. 62 (1999) 708 [arXiv:gr-qc/0002029]; Phys. Atom. Nucl. 62 (1999) 1524 [arXiv:gr-qc/0002030]; V.E. Kuzmichev and V.V. Kuzmichev, Eur. Phys. J. C 23 (2002) 337 [arXiv:astro-ph/0111438].

\bibitem{Ku2} V.E. Kuzmichev and V.V. Kuzmichev, Acta Phys. Pol. B 39 (2008) 979 [arXiv:0712.0464 [gr-qc]]; ibid. B 39 (2008) 2003 [arXiv:0712.0465 [gr-qc]]; ibid. B 40 (2009) 2877 [arXiv:0905.4142 [gr-qc]]; Ukr. J. Phys. 55 (2010) 626.

\bibitem{Mes} D. Meschini, Found. Sci. 12 (2007) 277 [arXiv:gr-qc/0601097]; 
C. Kiefer and B. Sandhoefer, in: Beyond the Big Bang, ed. R. Vaas, Springer, Heidelberg, 2008 [arXiv:0804.0672 [gr-qc]].

\bibitem{Lun} F. Lund, Phys. Rev. D 8 (1973) 3247;
V.G. Lapchinskii and V.A. Rubakov, Theor. Math. Phys. 33 (1977) 1076; 
F.J. Tipler, Rep. Prog. Phys. 68 (2005) 897.

\bibitem{Whi} G.J. Whitrow and D.G. Randall, MNRAS 111 (1951) 455.

\bibitem{Tol} R.C. Tolman, Relativity, Thermodynamics and Cosmology, Clarendon Press, Oxford, 1969, \S 139.

\bibitem{Di1} P.A.M. Dirac, Nature 139 (1937) 323.

\bibitem{Di2} P.A.M. Dirac, Pro. Roy. Soc. London A333 (1973) 403.

\bibitem{Bra} C.H. Brans and R.H. Dicke, Phys. Rev. 124 (1961) 925.

\bibitem{Wei} S. Weinberg, Gravitation and Cosmology, Wiley, New York, 1972.

\bibitem{Pet} J.P. Petit, Mod. Phys. Lett. A 3 (1988) 1527; J.W. Moffat, Int. J. Mod. Phys. D 2 (1993) 351 [arXiv:gr-qc/9211020]; A. Albrecht and J. Magueijo, Phys. Rev. D 59 (1999) 043516 [arXiv:astro-ph/9811018].

\bibitem{Lim} I. Prigogine, J. Geheniau, E. Gunzig and P. Nardone, Gen. Relativ.
Gravit. 21 (1989) 767; J.A.S. Lima, M.O. Calvao, and I. Waga, in: Cosmology, thermodynamics and matter creation in frontier physics, essays in honor of Jaime Tiomno, eds. S. MacDowel, H.M. Nussenzweig, R.A. Salmeron, World Scientific, Singapore, 1990 [arXiv:0708.3397 [astro-ph]]; M.O. Calvao, J.A.S. Lima, and I. Waga, Phys. Lett. A 162 (1992) 233; J.A.S. Lima and J. S.Alcaniz, Astron. Astrophys. 348 (1999) 1 [arXiv:astro-ph/9902337]; 
J.A.S. Lima, J.F. Jesus, and F.A. Oliveira, arXiv:1012.5069 [astro-ph.CO]; A. de Roany and J.A. de Freitas Pacheco, Gen. Relativ. Gravit. 43 (2011) 61.

\bibitem{Whe} J.A. Wheeler, in: Gravitation and Relativity, eds. Hong-Yee Chiu, W.F. Hoffmann, Benjamin, New York, 1964.

\bibitem{Hel} M. Heller, Acta Phys. Pol. B 1 (1970) 123.

\bibitem{Dym} I.G. Dymnikova and M.L. Fil'chenkov, Phys. Lett. B 545 (2002) 214 [arXiv:gr-qc/0209065].

\bibitem{Lar} D. Larson et al., Astrophys. J. Suppl. 192 (2011) 16 [arXiv:1001.4635 [astro-ph.CO]];
N. Jarosik et al., Astrophys. J. Suppl. 192 (2011) 14 [arXiv:1001.4744 [astro-ph.CO]].

\bibitem{Rie} A.G. Riess et al., Astrophys. J. 730 (2011) 119 [arXiv:1103.2976 [astro-ph.CO]].

\bibitem{Per} S. Perlmutter et al., Astrophys. J. 517 (1999) 565 [arXiv:astro-ph/9812133].

\bibitem{Ku3} V.V. Kuzmichev and V.E. Kuzmichev, Ukr. J. Phys. 50 (2005) 1321 [arXiv:astro-ph/0510763].

\end{thebibliography}
\end{document}